\documentclass[aps,prd,twocolumn,superscriptaddress]{revtex4}
\usepackage{epsfig,epsf}
\usepackage{amsmath}
\usepackage{amsthm}
\usepackage{amsfonts}
\usepackage{amssymb}
\usepackage{dsfont}
\usepackage{multirow}
\usepackage{appendix}
\usepackage{slashed}
\usepackage[active]{srcltx}
\usepackage{psfrag}
\usepackage{bbold}

\begin{document}

\title{ Possible Molecular Pentaquark States with Different Spin and Quark Configurations}

\date{\today}
\author{K.~Azizi}
\affiliation{Physics Department, Do\u gu\c s University,
Ac{\i}badem-Kad{\i}k\"oy, 34722 Istanbul, Turkey}
\affiliation{School of Physics, Institute for Research in Fundamental Sciences (IPM),
P.~O.~Box 19395-5531, Tehran, Iran}
\author{Y.~Sarac}
\affiliation{Electrical and Electronics Engineering Department,
Atilim University, 06836 Ankara, Turkey}
\author{H.~Sundu}
\affiliation{Department of Physics, Kocaeli University, 41380 Izmit, Turkey}

\begin{abstract}
We investigate three possible pentaquark candidates, one of which contains a single charm quark and the other two contain triple charm quarks in their  substructure. To this end we apply QCD sum rule method and  take into account both the positive and negative parity states corresponding to each possible pentaquark channel  having spin $3/2$ or $1/2$. Insisting on the importance of identification of the members of pentaquark family we obtain their spectroscopic parameters such as masses and residues. These parameters  are the main inputs  in the searches for their electromagnetic, strong and weak interactions.

\end{abstract}

\maketitle

\section{Introduction}

The exotic hadrons with non-conventional quark substructures have been investigated for many years. Having such  non-conventional configurations, different from the standard hadrons composed of tree quarks or a quark and an antiquark, make them interesting both theoretically and experimentally. Indeed, they have been searched for very long time in the experiment and their nature and probable internal structure have been theoretically investigated for many years. Finally the long sought result have been achieved and in 2003 $X(3872)$ was observed by Belle Collaboration~\cite{Choi:2003ue}. This  triggered the subsequent experimental searches to identify those non-conventional hadrons, especially $XYZ$ states, and measure their parameters. And finally the LHCb Collaboration~\cite{Aaij:2015tga} heralded the observation of another ones which are the  pentaquark states, $ P_c^+(4380) $ and $ P_c^+(4450) $. These states were reported to have possibly $ J^P=(3/2^-,5/2^+) $ quantum numbers, though not being well determined yet. These observations have triggered other investigations on such states and some other states were also interpreted as possible pentaquark states such as some of the newly observed $\Omega_c$ states by LHCb~\cite{Aaij:2017nav} as stated in Refs.~\cite{Kim:2017jpx} and, the states $N(1875)$ and $N(2100)$~\cite{He:2017aps}.  

We have a lack of knowledge about the inner structure and properties of these pentaquark states. To identify their structure different models were suggested. Among these models are the diquark-diquark antiquark model ~\cite{Wang:2016dzu,Anisovich:2015cia,Maiani:2015vwa,Ghosh:2015ksa,Wang:2015ava,Wang:2015epa,Wang:2015ixb,Ozdem:2018qeh}, diquark-triquark model~\cite{Wang:2016dzu,Zhu:2015bba,Lebed:2015tna}, topological soliton model~\cite{Scoccola:2015nia} and meson baryon molecular model~\cite{Wang:2016dzu,Ozdem:2018qeh,Yang:2015bmv,Burns:2015dwa,Lu:2016nnt,Tazimi:2016hsv,Shen:2016tzq,Guo:2013xga,Roca:2015dva,Chen:2015loa,Huang:2015uda,Meissner:2015mza,Xiao:2015fia,
He:2015cea,Chen:2015moa,Wang:2015qlf,Yamaguchi:2016ote,He:2016pfa,Azizi:2016dhy,Chen:2016heh,Wang:2017smo,Azizi:2018bdv}. Beside the observed $ P_c^+(4380) $ and $ P_c^+(4450) $ states  there are other possible candidates with possible five quark structure such as the ones studied in Ref~\cite{Chen:2016heh} in which the masses of charmed-strange molecular pentaquark states as well as other hidden charmed molecular ones were predicted. In Refs.~\cite{Chen:2016heh,Shimizu:2016rrd,Liu:2017xzo,yeni2,Azizi:2017bgs} along with the observed ones the pentaquak states containing $b$ quark were also investigated.    

The observation of pentaquark states by LHCb has brought some questions. One of them is about what possible internal structure these particles may have and whether they are tightly bound states or molecular ones. The other one is about the existence of the other possible stable pentaquark states. To shed light on these questions there have been an intense theoretical studies on these particles so far. However to understand them better, to identify their internal structure and their possible other candidates we need more investigations both on their spectroscopic properties and decay mechanism. The theoretical studies on these states may provide a deeper understanding on the nature and substructure of them and possible insights to the experimental researches as well as a deeper understanding on the strong interaction. With these motivations, in this work, we predict masses and residues of the three possible pentaquark states considering them in the meson-baryon molecular structure. For the investigation of the masses of these exotic particles we apply the QCD sum rules method~\cite{Shifman:1978bx,Shifman:1978by}. This method is among the effective nonperturbative methods which has been used widely in hadron physics giving reliable results consistent with the experimental observations.  

In this work we firstly consider recent announcement of the LHCb Collaboration on the observation of five new $ \Omega_c $ states in $ \Xi_c^+K^- $ channel~\cite{Aaij:2017nav}. In Refs~\cite{Chen:2017xat,Kim:2017jpx,Yang:2017rpg,Huang:2017dwn,Montana:2017kjw,Debastiani:2017ewu}  considering the closeness of their masses to a meson and a baryon threshold $ \Omega_c $ mesons were investigated with the possible molecular pentaquark assumption. Considering these interpretations we make a prediction on the mass of the possible molecular pentaquark states having single charm quark with spin parity $ J^P=\frac{3}{2}^{\pm} $. To this end, we chose a current in $\Xi_c^* \bar{K} $ molecular form.

In addition to these states considering another observation of LHCb Collaboration on double-charm baryon $ \Xi_{cc}^{++} $~\cite{Aaij:2017ueg} we study the possible triple charmed pentaquak states and calculate the masses and residues of them for both positive and negative parity cases. The interpolating currents in the calculations are chosen  in the  $ \Xi_{cc}(3621)D^0$ and $ \Xi_{cc}(3621)D^{*0}$ molecular form  with spin parity quantum numbers $ J^P=\frac{1}{2}^- $ and $ J^P=\frac{3}{2}^-$, respectively. Such a molecular interpretation of the possible triple charmed pentaquark state was also considered in Ref.~\cite{Chen:2017jjn} in which via one-boson-exchange model two possible molecular pentaquark states were predicted.  

The outline of the article is as follows. In Section II we present the detailed QCD sum rules calculations for the single charmed molecular pentaquark  and triple charmed pentaquark states.  Section III is devoted to the numerical analysis of the results. Finally we summarize and discuss our results in section IV.

\section{QCD sum rules calculation}
The details of the calculations for the considered possible three types of pentaquark states are presented in this section. In the calculation there are three steps to obtain QCD sum rules and these steps start from the correlation function. The mentioned correlation function is written in terms of the interpolating currents of the considered states and has a general form 
\begin{equation}
\Pi _{(\mu\nu)}(p)=i\int d^{4}xe^{ip\cdot x}\langle 0|\mathcal{T}\{J_{(\mu)}(x)\bar{J}_{(\nu)}(0)\}|0\rangle.   
\label{eq:CorrFPc}
\end{equation}
In the first step the above correlation function is calculated in terms of hadronic degrees of freedom such as mass of the hadron, current coupling constant of the hadron etc. This side of calculation is represented as physical or phenomenological side. In the second step the same correlation function is calculated in terms of QCD degrees of freedom containing mass of quarks and quark gluon condensates and named as theoretical or QCD side. Final step requires a match between the result of mentioned two sides of calculations considering the coefficient of same Lorentz structure from both sides. For the improvement of the analysis Borel transformation is used to suppress the contribution coming from higher states and continuum together with quark hadron duality assumption. 

\subsection{Phenomenological Side}

In this side we treat the interpolating currents as operators to annihilate or create the hadrons. To calculate the physical side, a complete set of hadronic states having the same quantum numbers with the considered interpolating current are inserted into the correlation function. Then the integration over $x$ is performed. The results appear in terms of masses and current coupling constant of the considered states, i.e. in terms of hadronic degrees of freedom.

\subsection*{The single charmed pentaquark states with $ J= \frac{3}{2}$ }

To calculate the physical side of the single charmed pentaquark states we follow the above given steps and firstly calculate the correlation functions in terms of hadronic degrees of freedom. For that purpose we insert  complete sets of hadronic state having the same quantum numbers with the considered interpolating current into correlation function. The integral over $x$ gives us the following result:  
\begin{eqnarray}
\Pi_{ \mu \nu }^{\mathrm{Phys}}(p)&=&\frac{\langle 0|J_{\mu }|{\frac{3}{2}}^{+}(p)\rangle
\langle {\frac{3}{2}}^{+}(p)|\bar{J}_{\nu }|0\rangle }{m_{\frac{3}{2}^+}^{2}-p^{2}} \nonumber \\&+&\frac{\langle 0|J_{\mu }|{\frac{3}{2}}^{-}(p)\rangle
\langle {\frac{3}{2}}^{-}(p)|\bar{J}_{\nu }|0\rangle }{m_{\frac{3}{2}^-}^{2}-p^{2}}+\cdots,
\label{eq:phys3bol2}\end{eqnarray}
where $m_{\frac{3}{2}^{+}}$ and $m_{\frac{3}{2}^{-}}$ represent the masses of the positive and negative parity particles, respectively. The ellipsis corresponds to contributions of the higher states and continuum. Using the following matrix elements 
\begin{eqnarray}
\langle 0|J_{\mu }|\frac{3}{2}^+(p)\rangle &=&\lambda_{\frac{3}{2}^{+}} \gamma_5 u_{\mu}^{+}(p),\nonumber\\
\langle 0|J_{\mu }|\frac{3}{2}^{-}(p)\rangle &=&\lambda_{\frac{3}{2}^{-}} u_{\mu}^{-}(p)
\label{eq:ResPcsingle}
\end{eqnarray}
parameterized in terms of the residues $\lambda_{\frac{3}{2}^+}$ and  $ \lambda_{\frac{3}{2}^-} $, and corresponding spinor, in Eq.~(\ref{eq:phys3bol2}) we obtain the Borel transformed correlation function as
\begin{eqnarray}
&&\mathcal{B}_{p^{2}}\Pi_{\mu \nu  }^{\mathrm{Phys}}(p)=-\lambda_{\frac{3}{2}^{+}}^{2}e^{-\frac{m_{\frac{3}{2}^{+}}^2}{M^2}}(-\gamma_5)({\slashed p}+m_{\frac{3}{2}^{+}})
  g_{\mu\nu}\gamma_5\nonumber \\
&&-
\lambda_{\frac{3}{2}^{-}}^{2}e^{-\frac{m_{\frac{3}{2}^{-}}^2}{M^2}}({\slashed p}+m_{\frac{3}{2}^{-}})
  g_{\mu\nu} +\cdots,
\nonumber \\  \label{eq:CorBorsingle3over2}
\end{eqnarray}
 where $ M^2 $ is the Borel mass squared.

\subsection*{The triple charmed pentaquark states with $ J= \frac{1}{2}$ and $ J= \frac{3}{2}$ }

Following similar steps as in single charmed case, we again start the calculation of the correlation functions in terms of hadronic degrees of freedom for triple charmed pentaquark states. Insertion of  complete sets of hadronic state and integration over $x$ gives us the following result:  
\begin{eqnarray}
\Pi^{\mathrm{Phys}}(p)&=&\frac{\langle 0|J|{\frac{1}{2}}^{+}(p)\rangle
\langle {\frac{1}{2}}^{+}(p)|\bar{J}|0\rangle }{m_{\frac{1}{2}^+}^{2}-p^{2}} \nonumber \\&+&\frac{\langle 0|J|{\frac{1}{2}}^{-}(p)\rangle
\langle {\frac{1}{2}}^{-}(p)|\bar{J}|0\rangle }{m_{\frac{1}{2}^-}^{2}-p^{2}}+\cdots,
\label{eq:physPccc}\end{eqnarray}
for spin$-1/2$ states, with masses $m_{\frac{1}{2}^{+}}$ and $m_{\frac{1}{2}^{-}}$ corresponding to the positive and negative parity particles, respectively. The ellipsis is again used for the representation of the contributions coming from the higher states and continuum. Using the following matrix elements 
\begin{eqnarray}
\langle 0|J|\frac{1}{2}^+(p)\rangle &=&\lambda_{\frac{1}{2}^{+}} \gamma_5 u(p),\nonumber\\
\langle 0|J|\frac{1}{2}^{-}(p)\rangle &=&\lambda_{\frac{1}{2}^{-}} u(p)
\label{eq:ResPccc}
\end{eqnarray}
in Eq.~(\ref{eq:physPccc}) the Borel transformed correlation function for this case is obtained as
\begin{eqnarray}
&&\mathcal{B}_{p^{2}}\Pi^{\mathrm{Phys}}(p)=-\lambda_{\frac{1}{2}^{+}}^{2}e^{-\frac{m_{\frac{1}{2}^{+}}^2}{M^2}}(-\gamma_5)({\slashed p}+m_{\frac{1}{2}^{+}})
  \gamma_5\nonumber \\
&&-
\lambda_{\frac{1}{2}^{-}}^{2}e^{-\frac{m_{\frac{1}{2}^{-}}^2}{M^2}}({\slashed p}+m_{\frac{1}{2}^{-}})
  +\cdots.
\nonumber \\  \label{eq:CorBorPccc}
\end{eqnarray}

As for the triple charmed states with spin$-3/2$ a similar procedure and similar steps as in single charmed pentaquark case are applied. Therefore we will skip the details for this calculation and remark that the results obtained here have the same forms as Eq.~(\ref{eq:phys3bol2}), Eq.~(\ref{eq:ResPcsingle}) and Eq.~(\ref{eq:CorBorsingle3over2}).  

Here we need to mention that for spin$-3/2$ parts, for both the single charmed and triple charmed pentaquark states, only the structures seen in Eq.~(\ref{eq:CorBorsingle3over2}) are given explicitly among the others. This is because of the fact that, these ones are the structures isolated from the spin$-1/2$ pollution and giving contributions to only spin$-3/2$ particles.

\subsection{Theoretical Side}

The second step in the QCD sum rule calculation requires the computation of the correlation function in terms of QCD degrees of freedom . In this part, the correlation function is reconsidered and it is calculated with explicit form of the interpolating currents of the interested states. In the calculations the quark fields present in interpolating currents are contracted via the Wick's theorem  which ends up with emergence of the light and heavy quark propagators. These quark  propagators are presented in~\cite{Azizi:2016dhy} in coordinate space and are used in the calculations, following which we transform the calculations to the momentum space by means of Fourier transformation. As in physical side, for the suppression of contribution of higher states and continuum we apply Borel transformation to this side also. Taking the imaginary parts of the results of the specified structure to be used in analysis we achieve the spectral densities.

\subsection*{The single charmed pentaquark states with $ J= \frac{3}{2}$ }

The interpolating current to be used in Eq.~(\ref{eq:CorrFPc})  for single charmed pentaquark states with spin$-3/2$ has the following form:
\begin{eqnarray}
J_{\mu}&=&[\epsilon^{abc}(q_{a}^{T}C\gamma_{\mu}s_{b})c_{c}][\bar{d}_{d}\gamma_{5}s_{d}].
 \label{eq:intJJPc}
\end{eqnarray}
In  Eq.~(\ref{eq:intJJPc}), the subscripts $a$, $b$, $c$ and $d$ are used to represent the color indices, $C$ is charge conjugation operator and $q$ represents $u$ or $d$ quark. This current does not only couple to the negative parity state but also to positive parity one. The reason for this can be explained as follows; multiplication of the current given in Eq.~(\ref{eq:intJJPc}) by $i\gamma_5$ gives a current $i\gamma_5 J_{\mu}$. This new form of the current will have opposite parity with respect to the current $J_{\mu}$. However, the calculations which are done by the new form of the current will not result in any new sum rules that are independent from the one that is done by the current $J_{\mu}$. Therefore the present calculations include the information of both parities. For more details on this subject one can see the Refs.~\cite{Wang:2015ava,Wang:2015epa,Wang:2015ixb,Chung,Bagan,Jido,Oka:1996zz}. In the present analysis we consider both the negative and the positive parity cases coupled to the current under consideration. Here we should also remark that the molecular type currents used in the present study also couple to the S-wave/P-wave  meson and baryon scattering states with the same quantum numbers and quark contents as the molecular pentaquark states under consideration.  Such contributions, which are entered to the physical sides of the calculations, have been taken into account for many multiquark systems in Refs.~\cite{Kondo:2004cr,Lee:2004xk,Sarac:2005fn,Matheus:2009vq}. However, in these studies it is found that the contributions of the meson and baryon scattering states  in multiquark systems are very small compared to the molecular pole contributions. For this, we ignore from such contributions in the present study.

Following the mentioned procedure, usage of interpolating current of single charmed state in correlation function and application of Wick's theorem results in   
\begin{eqnarray}
\Pi^{\mathrm{QCD}}_{\mu\nu}(p)&=& i\int d^{4}xe^{ip\cdot x}\epsilon^{abc}\epsilon^{a'b'c'}\Big\{\text{Tr}\left[\gamma_5 S_s^{db'}(x)\gamma_{\nu}\right.  \nonumber\\
&\times & \left. CS_q^{T{aa'}}(x)C\gamma_{\mu}S_s^{bd'}(x)\gamma_5 S_d^{d'd}(-x)\right] S_c^{cc'}(x)\nonumber\\
&-&\text{Tr}\left[\gamma_5 S_s^{dd'}(x)\gamma_{5} S_d^{d'd}(-x)\right] \text{Tr}\left[\gamma_{\nu}C S_q^{T{aa'}}(x)\right.\nonumber\\
&\times &\left. C\gamma_{\mu} S_s^{bb'}(x)\right] S_c^{cc'}(x)\Big\}.  \label{eq:CorrF1Pc}
\end{eqnarray}
Then the propagators of light and heavy quarks are used in this equation and  following straightforward mathematical calculations we obtain the results for this side. Imaginary parts of the results obtained for chosen Lorentz structures  provide us with the spectral densities. To provide  samples for the spectral densities obtained in this work, we present the results of this subsection in the Appendix.  

\subsection*{The triple charmed pentaquark states with $ J= \frac{1}{2}$ and $ J= \frac{3}{2}$ }

The interpolating currents used for triple charm pentaquark states with spin $ J= \frac{1}{2}$ and $ J= \frac{3}{2}$ are as follows:
\begin{eqnarray}
J&=&[\epsilon^{abc}(c_{a}^{T}C\gamma_{\mu}c_{b})\gamma^{\mu}\gamma_{5}q_{c}][\bar{u}_{d}\gamma_{5}c_{d}],\nonumber\\
J_{\mu}&=&[\epsilon^{abc}(c_{a}^{T}C\gamma_{\theta}c_{b})\gamma^{\theta}\gamma_{5}q_{c}][\bar{u}_{d}\gamma_{\mu}c_{d}],
 \label{eq:JJPccc}
\end{eqnarray}
respectively. These currents also couple to both the positive and negative parity states from similar reason  stated in the previous case The results for the triple charmed states are obtained after the contraction as 
\begin{eqnarray}
\Pi^{\mathrm{QCD}}(p)&=&\mp i\int d^{4}xe^{ip\cdot x}\epsilon^{abc}\epsilon^{a'b'c'} \gamma_{\mu}\gamma_5 S_q^{cc'}(x)\gamma_5 \gamma_{\nu} \nonumber\\
&\times& \Big\{ \text{Tr}\left[\gamma_{\nu} C S_c^{T{bb'}}(x)C\gamma_{\mu}S_c^{ad'}(x)\gamma_{i}S_u^{d'd}(-x)\right.\nonumber\\
&\times& \left. \gamma_{j}S_c^{da'}(x)\right]-\text{Tr}\left[\gamma_{\nu} C S_c^{T{ba'}}(x)C\gamma_{\mu}S_c^{ad'}(x)\right.\nonumber\\
&\times& \left.\gamma_{i}S_u^{d'd}(-x) \gamma_{j}S_c^{db'}(x)\right]+\text{Tr}\left[\gamma_{\nu} C S_c^{T{ab'}}(x)\right.\nonumber\\&\times& \left.C\gamma_{\mu}S_c^{bd'}(x)\gamma_{i}S_u^{d'd}(-x) \gamma_{j}S_c^{da'}(x)\right]\nonumber\\&-&\text{Tr}\left[\gamma_{\nu} C S_c^{T{aa'}}(x)C\gamma_{\mu}S_c^{bd'}(x)\gamma_{i}S_u^{d'd}(-x)\right.\nonumber\\
&\times& \left. \gamma_{j}S_c^{dd'}(x)\right]+\text{Tr}\left[\gamma_{\nu} C S_c^{T{bb'}}(x)C\gamma_{\mu}\right.\nonumber\\&\times&\left. S_c^{aa'}(x)\right]\text{Tr}\left[\gamma_{i}  S_u^{d'd}(-x)\gamma_{j}S_c^{dd'}(x)\right]\nonumber\\&-&
\text{Tr}\left[\gamma_{\nu} C S_c^{T{ba'}}(x)C\gamma_{\mu}S_c^{ab'}(x)\right]\nonumber\\&\times& \text{Tr}\left[\gamma_{i}  S_u^{d'd}(-x)\gamma_{j}S_c^{dd'}(x)\right]\Big\}.
\label{eq:CorrF1Pccc}
\end{eqnarray}
In Eq.~(\ref{eq:CorrF1Pccc}) the $-$ and $+$ signs at the beginning of the equation are for spin$-1/2$ and spin$-3/2$ particles, respectively and the $\gamma_i$ and $\gamma_j$ is used for $\gamma_i=\gamma_j=\gamma_5$ for spin$-1/2$ and $\gamma_i=\gamma_{\alpha'}$ and $\gamma_j=\gamma_{\alpha}$ for spin$-3/2$ case, respectively.

\subsection{QCD sum rules }

After the calculations of both sides are completed we choose the same Lorentz structures from each side and we match the coefficients to obtain the QCD sum rules giving us the physical quantities that we seek for. From this procedures we obtain 
\begin{eqnarray}
&&m_{i^+}\lambda_{i^+}^{2}e^{-m_{i^+}^{2}/M^{2}}- m_{i^-}\lambda_{i^-}^{2}e^{-m_{i^-}^{2}/M^{2}}=%
\Pi^{m}_{i},\nonumber\\
&&\lambda_{i^+}^{2}e^{-m_{i^+}^{2}/M^{2}}+\lambda_{i^-}^{2}e^{-m_{i^-}^{2}/M^{2}}=%
j\Pi^{p}_{i},
\label{eq:srcoupling1}
\end{eqnarray}
for single and triple charmed pentaquark states, where  $i^{\pm}$ are used to represent the spin$-1/2^{\pm}$ and spin-$3/2^{\pm}$ states. $j$ is $+$ for spin-$1/2$ and $-$ for spin-$3/2$ cases. The $\Pi^{m}_{i}$ and $\Pi^{p}_{i}$, which are the same for both the positive and negative parities in the corresponding channel,  are the functions respectively obtained in QCD side from the coefficients of the structures $\mathbb{1}$ and $ {\slashed p}$ for spin$-1/2$ and $ g_{\mu\nu} $ and $ {\slashed p} g_{\mu\nu}$ for spin$-3/2$ cases and they are written as 
\begin{equation}
\Pi^{m(p)}_{i}=\int^{s_0}_{s'}ds\rho^{m(p)}_{i}(s)e^{-s/M2},
\label{spectral}
\end{equation}%
in terms of spectral densities, where $ s_0 $ is the continuum threshold, $s'=(2m_s+m_c)^2$ for single charmed pentaquark and $s'=9m_c^2$ for triple charmed ones.  The spectral densities $\rho^{m(p)}$ contain both perturbative and nonperturbative parts and can be represented for each structure denoted by $m(p)$ as
\begin{equation}
\rho^{m(p)}_{i} (s)=\rho_{i} ^{m(p),\mathrm{pert.}}(s)+\sum_{k=3}^{6}\rho^{m(p)}_{i,k}(s),
\label{eq:A1}
\end{equation}%
with  $\sum_{k=3}^{6}\rho^{m(p)}_{i,k}(s)$ part containing  the   nonperturbative contributions of dimensions three, four, five and six . In the Appendix we present the results of spectral densities obtained for the single charmed pentaquark state to provide an example.  

To obtain the present four unknown physical quantities, namely $\lambda_{i^+}$, $\lambda_{i^-}$, $m_{i^+}$ and $m_{i^-}$  for each possible pentaquark state considered in this work, beside the two equations given in Eq.~(\ref{eq:srcoupling1})  we need two more equations. We obtain them taking the derivative of both sides of Eq.~(\ref{eq:srcoupling1}) with respect to $\frac{1}{M^2}$. Simultaneous solution of the obtained four equations give the desired physical quantities in terms of the QCD degrees of freedom, continuum threshold and Borel parameter. Note that
the resultant equations are four nonlinear coupled equations that we will solve them numerically to find the four unknown quantities in next section.

\section{Numerical results}
The sum rules obtained in the last subsection contain QCD degrees of freedom, Borel parameter $ M^2 $ as well as continuum threshold $ s_0 $. These are all input parameters in the calculations to acquire the physical quantities of interest by numerical solving of the sum rules of four nonlinear coupled equations. Among these input parameters are the masses of light quarks $u$ and $d$ and they are taken as zero. Table~\ref{tab:Param} includes some of these input parameters.
\begin{table}[tbp]
\begin{tabular}{|c|c|}
\hline\hline
Parameters & Values \\ \hline\hline
$m_{c}$ & $(1.28\pm 0.03)~\mathrm{GeV}$ \\
$\langle \bar{q}q \rangle $ & $(-0.24\pm 0.01)^3$ $\mathrm{GeV}^3$  \\
$\langle \bar{s}s \rangle $ & $m_0^2\langle \bar{q}q \rangle$  \\
$m_{0}^2 $ & $(0.8\pm0.1)$ $\mathrm{GeV}^2$ \\
$\langle \overline{q}g_s\sigma Gq\rangle$ & $m_{0}^2\langle \bar{q}q \rangle$
\\
$\langle \overline{s}g_s\sigma Gs\rangle$ & $m_{0}^2\langle \bar{s}s \rangle$
\\
$\langle\frac{\alpha_sG^2}{\pi}\rangle $ & $(0.012\pm0.004)$ $~\mathrm{GeV}%
^4 $\\
\hline\hline
\end{tabular}%
\caption{Some input parameters used in the calculations.}
\label{tab:Param}
\end{table}

 In the analysis we have two auxiliary parameters: threshold parameter $s_0$ and Borel parameters $M^{2}$. To carry over the analysis their working intervals are needed. To determine these intervals one needs the criteria which bring some limitations on their values. For the Borel window these criteria are the convergence of the series of OPE and the adequate suppression of the contributions of higher states and continuum. To determine the lover limit of the interval of Borel parameter we consider the OPE  convergence and demand the contribution coming from the higher dimensional term in the OPE should be less than the others, in our case it constitute almost 4\% of the total OPE. As for the upper limit of this parameter, we consider the pole contribution to be greater  than the contributions of the higher states and continuum. We fix the maximum value of Borel parameter imposing the pole contribution to be greater or at least equal 50\% of the  total. The threshold parameter is  not completely arbitrary and it is related to the energy of the first corresponding  excited state. In its fixing we again consider the pole dominance and OPE convergence. To depict how  the OPE converge in our calculations  Fig.~\ref{fig:OPE} is presented. In this figure it can be easily  seen that the contributions coming from different operators decrease with the increasing the dimension and the perturbative one has the dominant contribution. And also to show the dominance of pole contribution we give  Fig.~\ref{fig:pole} which shows the ratio of the pole contribution to the total as
 \begin{equation}
 \mbox{PC}=\frac{\Pi(M^2,s_0)}{\Pi(M^2,\infty)}
 \end{equation}
 for the chosen intervals of auxiliary parameters. From this figure, we see that the pole contribution dominates over the contributions of the 
 higher states and continuum and constitutes the main part of the total contributions. 
\begin{widetext}

\begin{figure}[h!]
\begin{center}
\includegraphics[totalheight=5cm,width=7cm]{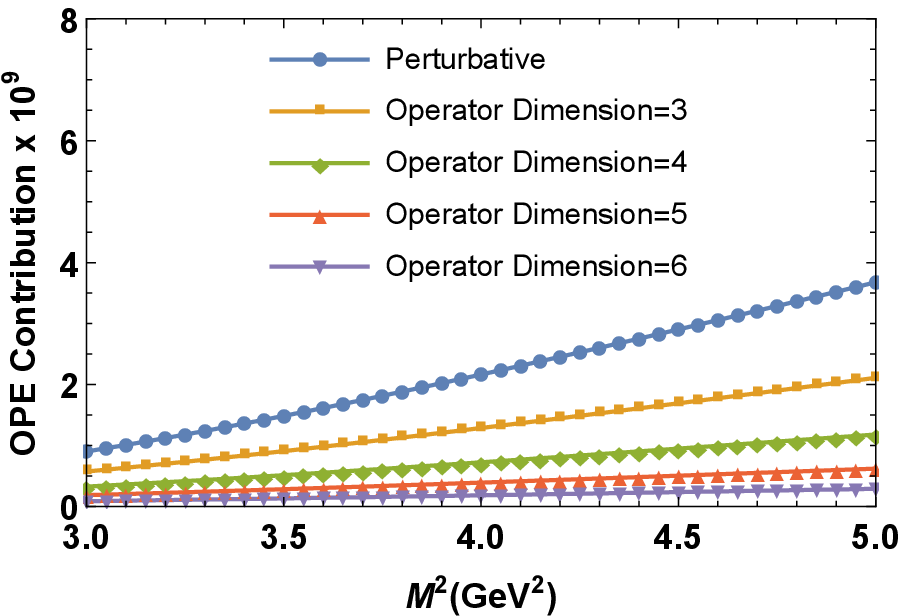}
\includegraphics[totalheight=5cm,width=7cm]{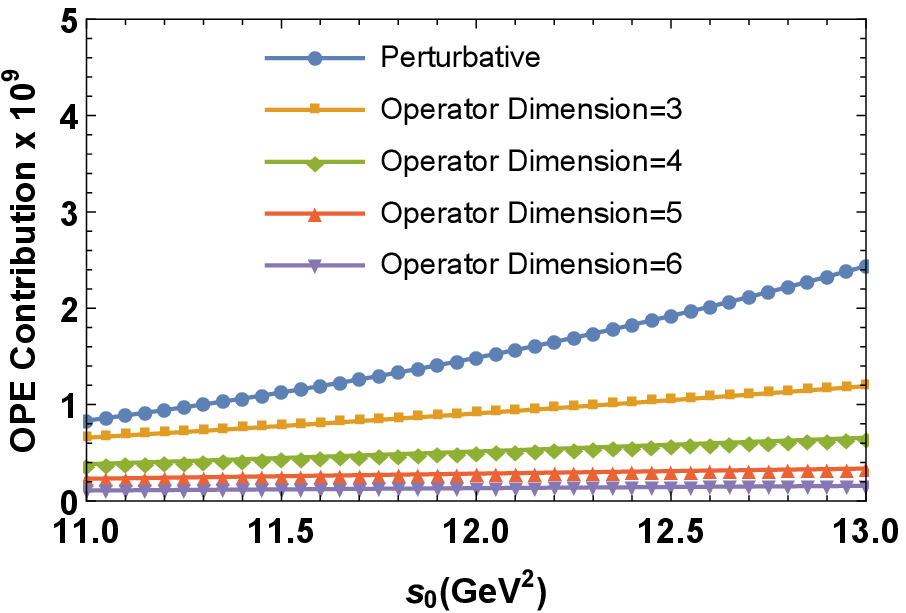}
\end{center}
\caption{\textbf{Left:} The OPE contribution for the possible $ \Xi_c^*\bar{K} $ molecular pentaquark as a function of Borel
parameter $M^2$  at the central value of the continuum threshold $s_0$. \textbf{Right:}
The OPE contribution for the possible $ \Xi_c^*\bar{K}$ molecular pentaquark as a function of threshold
parameter $s_0$ at the centrslvalue of the Borel parameter $M^2$. } \label{fig:OPE}
\end{figure}
\begin{figure}[h!]
\begin{center}
\includegraphics[totalheight=5cm,width=7cm]{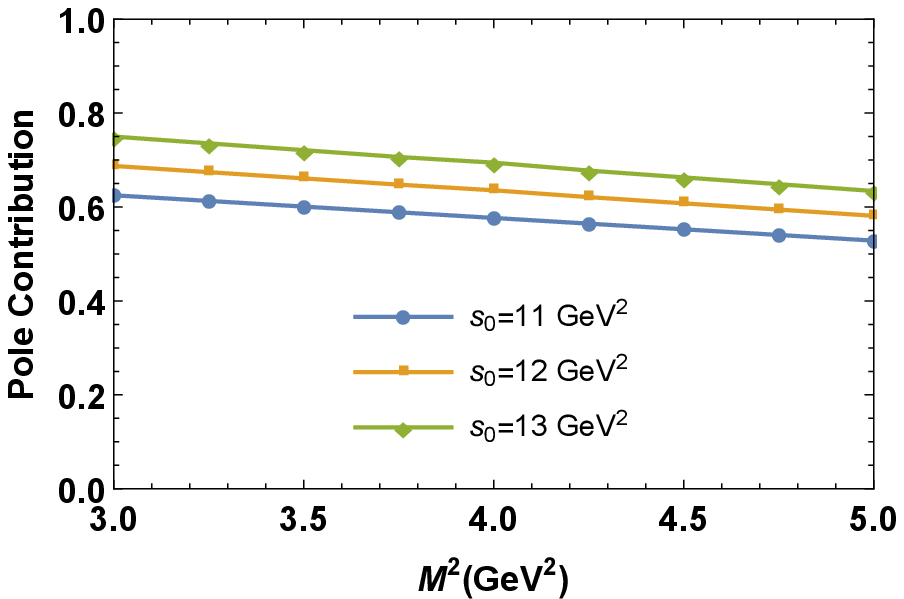}
\includegraphics[totalheight=5cm,width=7cm]{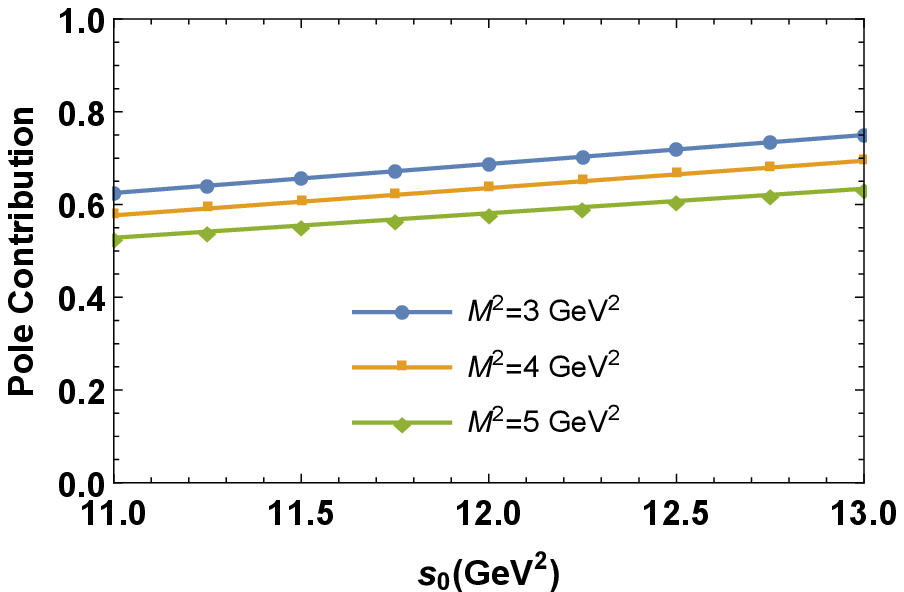}
\end{center}
\caption{\textbf{Left:} The pole contribution for the possible pentaquark having molecular form $ \Xi_c^*\bar{K} $ as a function of Borel
parameter $M^2$  at different fixed values of the continuum threshold $s_0$. \textbf{Right:}
The pole contribution for the possible pentaquark having molecular form $ \Xi_c^*\bar{K} $ as a function of the continuum threshold $s_0$ at different fixed values of the Borel parameter $M^2$. } \label{fig:pole}
\end{figure}
\end{widetext}
The analyses done with these criteria result in the intervals given in Table~\ref{tab:Auxuliaryparam} for these parameters:
\begin{table}[tbp]
\begin{tabular}{|c|c|c|c|}
\hline\hline
                       & $J^P$    & $M^2$ (GeV$^2$)& $s_0$ (GeV$^2$) \\ \hline\hline
$ \Xi_c^*\bar{K} $     & $3/2^+$  &  $3-5$         &$11-13$          \\ 
                       & $3/2^-$  &                &                 \\ \hline
$\Xi_{cc}(3621)D^0$    & $1/2^+$  &  $6-8$         &$40-42$          \\ 
                       & $1/2^-$  &                &                 \\ \hline
$\Xi_{cc}(3621)D^{*0}$ & $3/2^+$  &  $6-8$         & $40-42$         \\ 
                       & $3/2^-$  &                &                 \\ \hline\hline
\end{tabular}%
\caption{Working intervals of Borel masses $M^2$ and threshold parameters $s_0$ used in the calculations.}
\label{tab:Auxuliaryparam}
\end{table}

Now, as examples, we would like to draw the graphs for masses and residues of the positive and negative parity states pointing out the dependencies of the results obtained for $ \Xi_c^*\bar{K} $ molecular pentaquark on Borel mass $M^2$ and threshold parameter $s_0$ in figures \ref{mass32Msq}-\ref{residue32s0}. These graphs depict weak dependencies of the results on the auxiliary parameters in their working intervals  as it is  expected considering the good convergence of the OPE and sufficient pole contribution. Our analyses show that the dependencies of the results  on the auxiliary parameters in their working intervals are relatively weak compared to the regions out of these windows. The Borel parameter is a mathematical  object coming from the Borel transformation. Although no dependence on it is expected in reality, the relatively weak dependence is acceptable in practice bringing some uncertainty to the calculations. As we stated above, the continuum threshold is not totally arbitrary and it depends on the energy of the first excited state with the same quantum numbers as the interpolating currents. Hence, the relatively obvious dependencies of the results on this parameters are reasonable compared to the dependencies on the pure mathematical Borel parameter.  In the calculations, considering the standard prescriptions of the QCD sum rule method, suitable regions for the Borel mass $M^2$ and threshold parameter $s_0$ are chosen so that in these regions one gets the possible maximum stability for the mass and residue. The weak dependencies of the results shown in the figures on the auxiliary  parameters are acceptable in the QCD sum rule calculations since the obtained uncertainties remain inside the typical limits of the standard  error range of the QCD sum rule method not exceeding the 30\% of the total result. Besides, as mentioned above, the chosen regions for the auxiliary parameters provide us with a good OPE convergence and pole dominance required by the method to have reliable results. The uncertainties coming from the variations of the results with respect to the variations of the auxiliary parameters  manifest themselves as errors in the results. 
\begin{widetext}

\begin{figure}[h!]
\begin{center}
\includegraphics[totalheight=5cm,width=7cm]{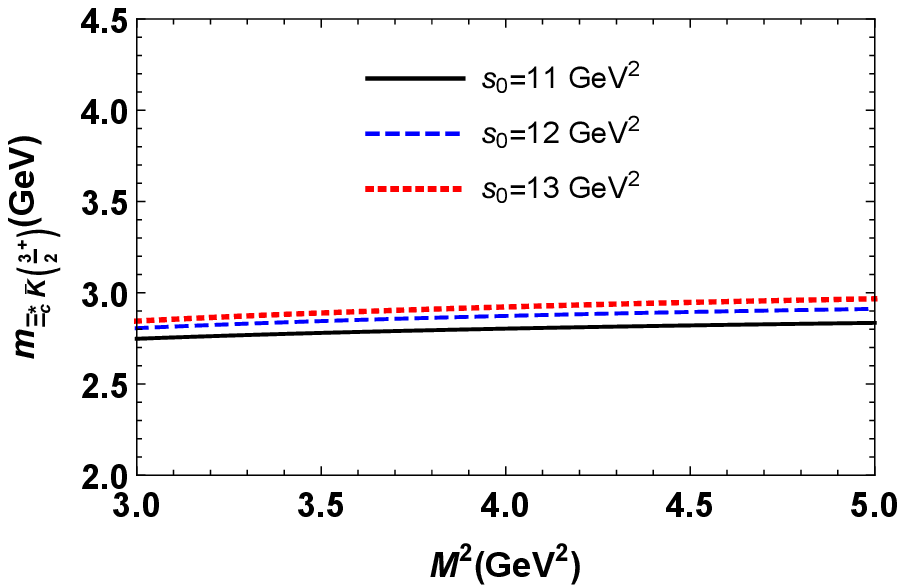}
\includegraphics[totalheight=5cm,width=7cm]{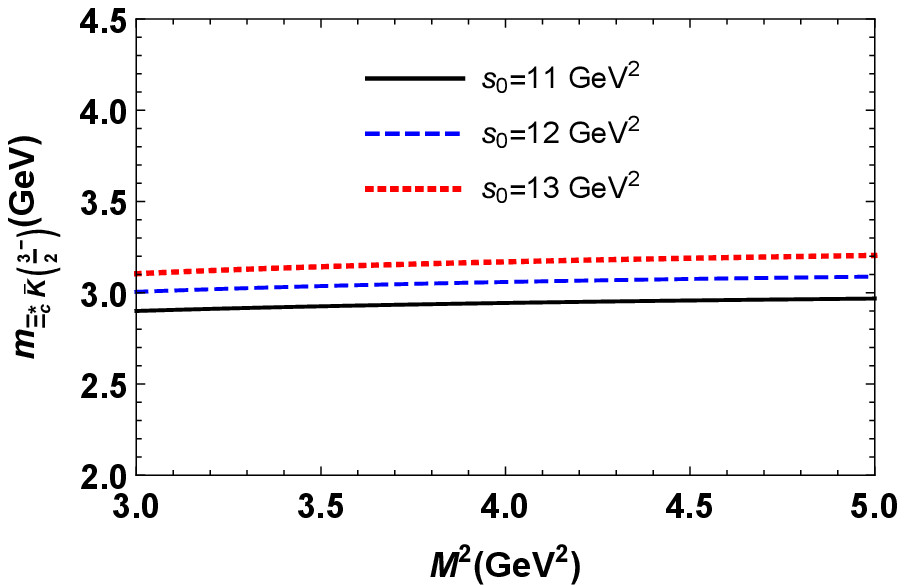}
\end{center}
\caption{\textbf{Left:} The mass of the possible pentaquark having molecular form $ \Xi_c^*\bar{K} $ with positive parity as a function of Borel
parameter $M^2$  at different fixed values of the continuum threshold. \textbf{Right:}
 The mass of the possible pentaquark having molecular form $ \Xi_c^*\bar{K} $ with negative parity as a function of Borel
parameter $M^2$  at different fixed values of the continuum threshold. } \label{mass32Msq}
\end{figure}

\begin{figure}[h!]
\begin{center}
\includegraphics[totalheight=5cm,width=7cm]{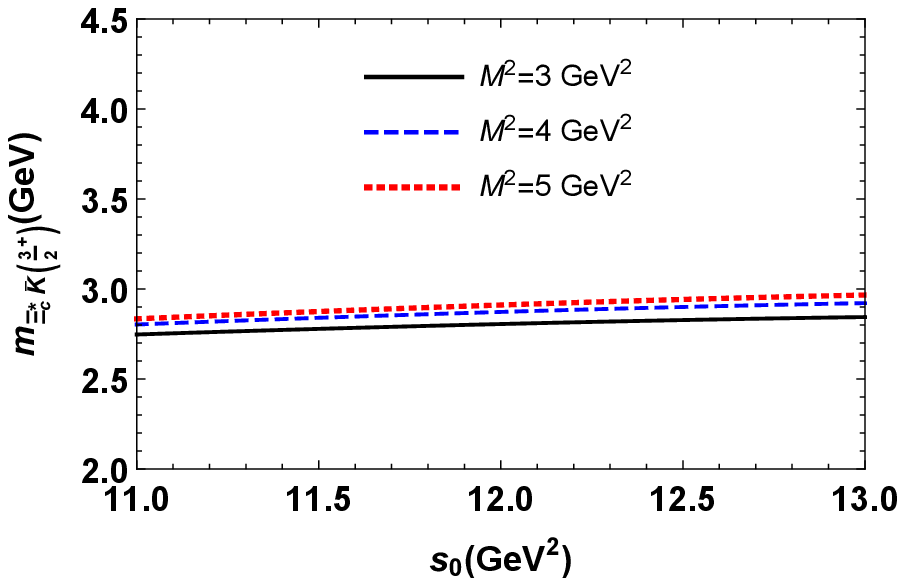}
\includegraphics[totalheight=5cm,width=7cm]{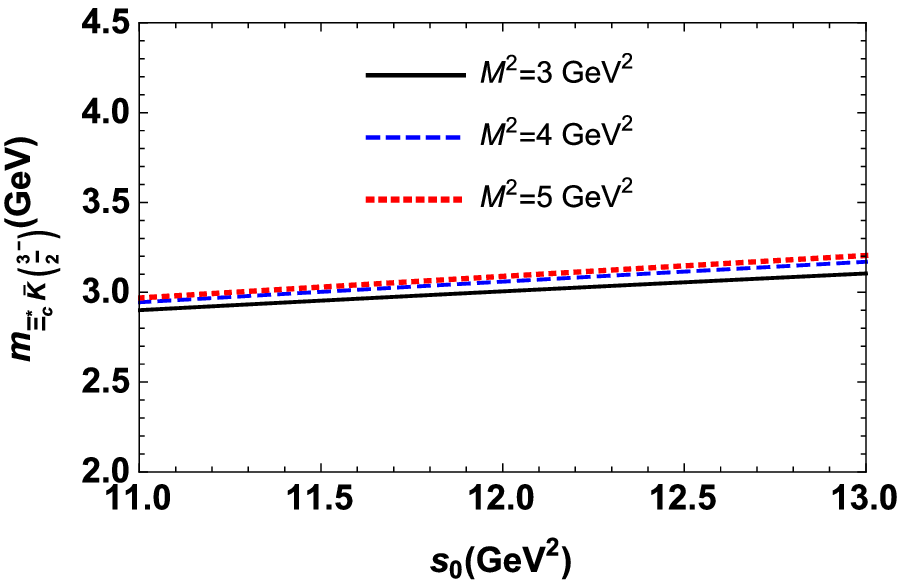}
\end{center}
\caption{\textbf{Left:} The mass of the possible pentaquark having molecular form $ \Xi_c^*\bar{K} $ with positive parity as a function of threshold
parameter $s_0$  at different fixed values of the Borel parameter. \textbf{Right:}
 The mass of the possible pentaquark having molecular form $ \Xi_c^*\bar{K} $ with negative parity as a function of Borel
parameter $s_0$   at different fixed values of the Borel parameter. } \label{mass32s0}
\end{figure}

\begin{figure}[h!]
\begin{center}
\includegraphics[totalheight=5cm,width=7cm]{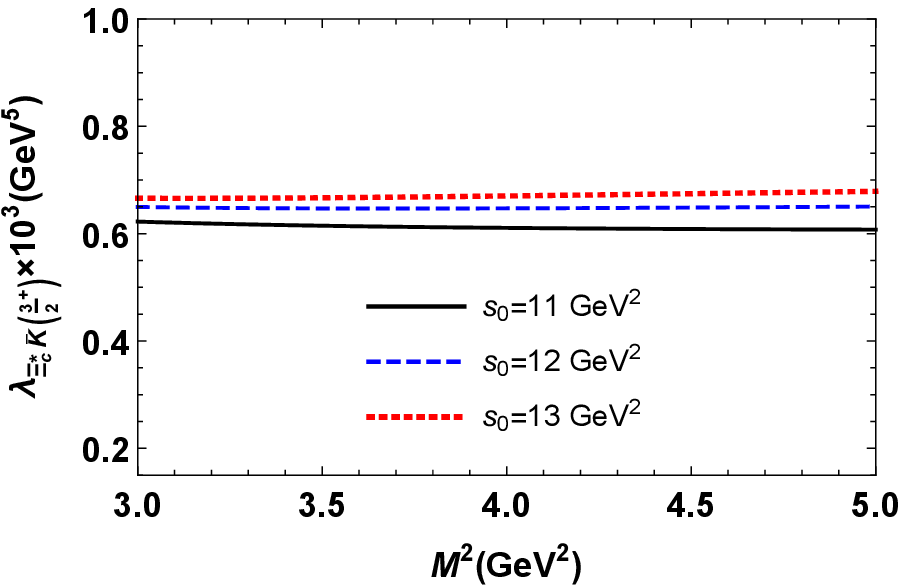}
\includegraphics[totalheight=5cm,width=7cm]{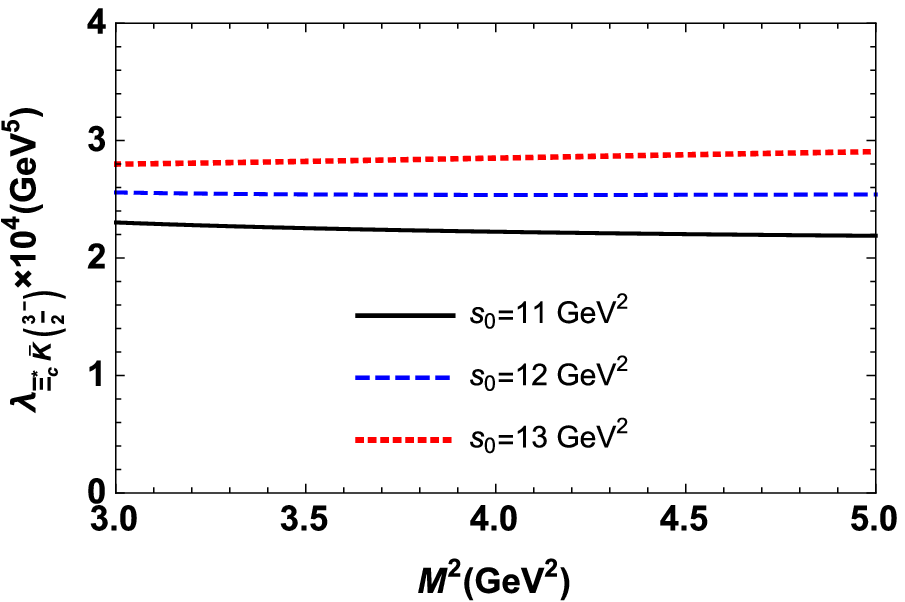}
\end{center}
\caption{\textbf{Left:} The residue of the possible pentaquark having molecular form $ \Xi_c^*\bar{K} $ with positive parity as a function of  $M^2$  at different fixed values  of the continuum threshold.
 \textbf{Right:}
 The residue of the possible pentaquark having molecular form $ \Xi_c^*\bar{K} $ with negative parity as a function of  $M^2$  at different fixed values of the continuum threshold.} \label{residue32Msq}
\end{figure}

\begin{figure}[h!]
\begin{center}
\includegraphics[totalheight=5cm,width=7cm]{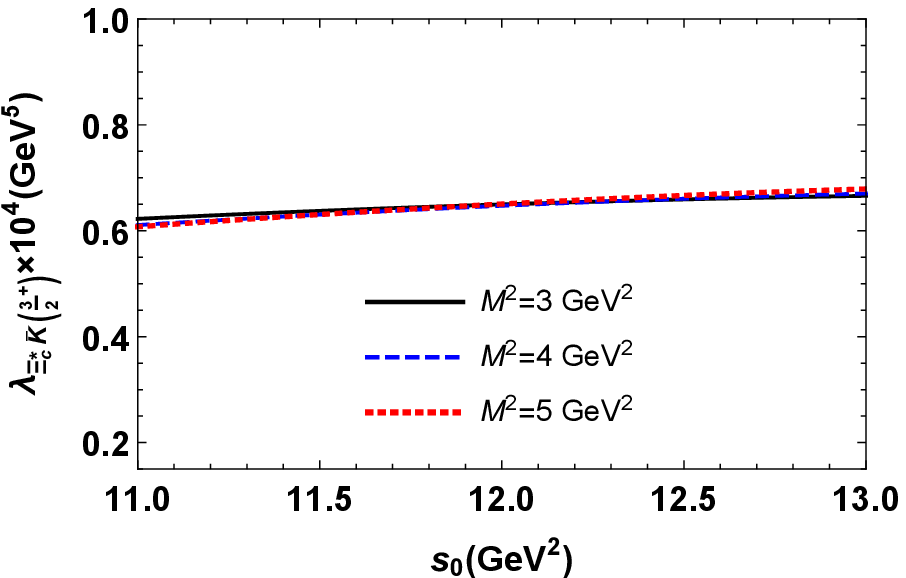}
\includegraphics[totalheight=5cm,width=7cm]{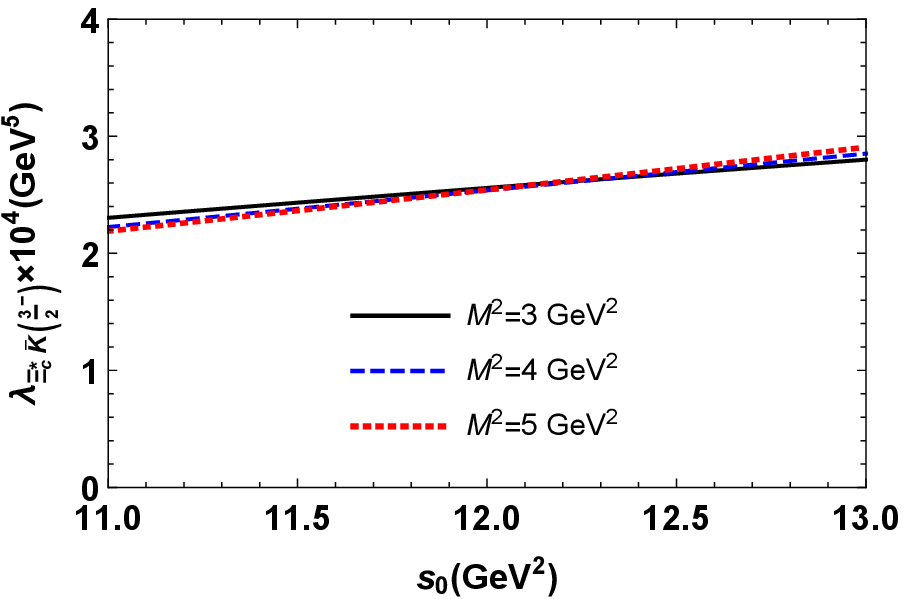}
\end{center}
\caption{\textbf{Left:} The residue of the possible pentaquark having molecular form $ \Xi_c^*\bar{K} $ with positive parity as a function of  $s_0$  at different fixed values  of the Borel parameter.
\textbf{Right:}
 The residue of the possible pentaquark having molecular form $ \Xi_c^*\bar{K} $ with positive parity as a function of  $s_0$   at different fixed values of the Borel parameter.} \label{residue32s0}
\end{figure}

\end{widetext}
The working intervals and the other input parameters are used in the QCD sum rule results to obtain the physical parameters  of the  states that we address. Table~\ref{tab:results} presents these results with their corresponding errors. The uncertainties arise due to the errors included in input parameters and those inherited from determination of the intervals of auxiliary parameters. 
\begin{table}[tbp]
\begin{tabular}{|c|c|c|c|}
\hline\hline
                      & $J^P$   & $m~(\mathrm{MeV})$& $\lambda~(\mathrm{GeV}^{5})$ \\ \hline\hline
$ \Xi_c^*\bar{K} $    & $3/2^+$ & $2856^{+55}_{-109}$   &$0.65^{+0.06}_{-0.03}\times 10^{-4}$  \\ 
                      & $3/2^-$ & $3049^{+155}_{-149}$ &$2.59^{+0.36}_{-0.36}\times 10^{-4}$    \\ \hline
$ \Xi_{cc}(3621)D^0$  & $1/2^+$ &$5601^{+148}_{-109}$  &$1.64^{+0.29}_{-0.28}\times 10^{-3}$ \\ 
                      & $1/2^-$ &$5583^{+209}_{-212}$  &$1.61^{+0.29}_{-0.27}\times 10^{-3}$\\  \hline
$\Xi_{cc}(3621)D^{*0}$& $3/2^+$ & $5726^{+167}_{-118}$ & $4.37^{+0.49}_{-0.43}\times 10^{-3}$\\ 
                      & $3/2^-$ & $5728^{+228}_{-279}$ & $4.58^{+0.56}_{-0.58}\times 10^{-3}$\\ 
\hline\hline
\end{tabular}%
\caption{The results of QCD sum rules calculations for the masses and residues of the possible pentaquark states.}
\label{tab:results}
\end{table}

A similar mass prediction on possible pentaquark state containing single charm quark was made in Ref.~\cite{Wang:2018alb} using QCD sum rule method. In this work a diquark-diquark-antiquark type current was considered and the result for the $J^P=3/2^-$ state was obtained as $3.15\pm 0.13$~GeV. Another prediction for possible single charmed pentaquark in diquark-diquark-antiquak model was presented in Ref.~\cite{Anisovich:2017aqa} and the estimation for the mass of $J^P=3/2^-$ state was given as $3.2\pm 0.1$~GeV. These result are consistent with ours within the errors. As for the triply charmed pentaquark state, the spin$-1/2$ case is studied in Ref.~\cite{Wang:2018ihk} in diquark-diquark-antiquark configuration and the corresponding masses and residues are given as $M=5.61\pm0.10$~GeV, $\lambda=(2.38\pm0.31 \times 10^{-3})$~GeV$^{5}$ and $M=5.72\pm0.10$~GeV, $\lambda=(1.45\pm0.28 \times 10^{-3})$~GeV$^{5}$ for negative and positive parities, respectively. These results are again in consistency with ours considering the error ranges. Looking at these results we may state that for such possible pentaquark states both the molecular and diquark-diquark-antiquark  interpretations can be considered  for their inner structures. Therefore to identify them we need more theoretical works not only on the spectroscopic properties of these type of particles but also on their possible interactions with other particles. On the other hand one can not look over the contribution of such theoretical studies for gaining  deeper understanding in the nonperturbative realm of QCD.

\section{Summary and Outlook}
In this work we consider some possible pentaquark states containing single or triple charm quark. We assign their structure in molecular form and find their masses and residues using QCD sum rules method. The calculations include both positive and negative parity states corresponding to each pentaquarks. The  single charmed pentaquark state is considered as $\Xi_c^*\bar{K}$ molecular state with $J^P=3/2^\pm$ and the triple charmed pentaquarks are as $ \Xi_{cc}(3621)D^0$ and $\Xi_{cc}(3621)D^{*0}$ molecular states with corresponding  $J^P=1/2^\pm$ and  $J^P=3/2^\pm$, respectively. The results obtained in this work are compared with the other present results for differently chosen quark configurations in literature. From this comparison it has been seen that the obtained results are in agreement. The results of present study may give an insight into the future experimental searches but it is clear that to distinguish the inner structure of prospective pentaquark states having such quark substructure these mass predictions, though necessary, may not be enough and it is needed to study other properties of them such as their possible decays. Hence, it is important to study such states theoretically in different respects not only to provide some insights into the future experiments but also to better understand the properties of these possible states. The theoretical studies on these states will also improve our knowledge on the present pentaquark states as well as on the nonperturbative nature of the QCD.

As final remark, we shall state that the interpolating currents used in the present study not only couple to the considered meson-baryon molecular pentaquark states but also to the meson and baryon scattering states with the same quantum numbers and quark contents. It was previously shown in Refs.~\cite{Kondo:2004cr,Lee:2004xk,Sarac:2005fn,Matheus:2009vq} that the contributions of the scattering states are very small compared to the molecular pole contributions in multiquark systems. Therefore, we ignored the meson and baryon scattering effects and our results are valid within this approximation.

\section*{ACKNOWLEDGEMENTS}

The authors thank  T\"{U}B\.{I}TAK for partial support provided under the Grant no: 115F183.

\label{sec:Num}

\section*{APPENDIX: SPECTRAL DENSITIES}

To exemplify the spectral density results, in this appendix, the perturbative and nonperturbative parts (with dimensions three, four, five and six) of the spectral densities for the single charmed pentaquark states are presented in terms of the Feynman parameters $x$ and $ y $. These results are corresponding to the coefficients of the structures $g_{\mu\nu}$  and  ${\slashed p} g_{\mu\nu}$.

\begin{widetext}

For the structure $g_{\mu\nu}$:

\begin{eqnarray}
\rho^{m,\mathrm{pert.}}_{\frac{3}{2}}&=&\int\limits_{0}^{1} dx \frac{m_cx^4(m_c^2+sr)^4\Big(30m_s^2r(-4+r)-11(m_c^2+sr)x(-5+r)\Big)}
{2^{20}\cdot5^2\cdot3^2\pi^8r^5} \Theta[L],\nonumber \\
\rho^{m}_{\frac{3}{2},3}&=&\int\limits_{0}^{1} dx \frac{m_cm_sx^3(m_c^2+sr)^3\Big(10\langle
\bar{d}d\rangle(-3+r)-40\langle \bar{q}q\rangle-13\langle
\bar{s}s\rangle(-3+r)\Big)}{2^{15}\cdot3^2\pi^6r^3}\Theta[L],
\nonumber \\
\rho^{m}_{\frac{3}{2},4}&=&-\int\limits_{0}^{1} dx \langle \frac{\alpha_s GG}{\pi}\rangle\frac{
x^2m_c(m_c^2+sr)}{5\cdot3^3\cdot2^{19}\pi^6r^4}\Big[5m_c^4x(180-263x+67x^2)+sr^2
\Big(sx(900-1315x+269x^2+11x^3)
\nonumber \\
&+&6m_s^2(30-5x^2-3x^3)\Big)+m_c^2r\Big(6m_s^2(30-15x^2-x^3)
+sx(1800-2630x+604x^2+11x^3)\Big)\Big]\Theta[L],
\nonumber \\
\rho^{m}_{\frac{3}{2},5}&=&\int\limits_{0}^{1} dx \frac{m_cm_sx^2(m_c^2+sr)^2  m_0^2\Big(45\langle
\bar{q}q\rangle-15\langle \bar{d}d\rangle(-2+r)+14\langle
\bar{s}s\rangle(-2+r)\Big)}{2^{14}\cdot3^2\pi^6r^2}\Theta[L],
\nonumber \\
\rho^{m}_{\frac{3}{2},6}&=&\int\limits_{0}^{1} dx \Bigg\{\frac{m_cx^2(m_c^2+sr)^2\Big[\langle
\bar{s}s\rangle\Big(30\langle \bar{q}q\rangle+\langle
\bar{s}s\rangle(-2+r)\Big)-\langle \bar{d}d\rangle\Big(3\langle
\bar{q}q\rangle+10\langle
\bar{s}s\rangle(-2+r)\Big)\Big]}{2^{11}\cdot3^2\pi^4r^2}
\nonumber \\
&-& \frac{11m_cx^2(-2+r)(m_c^2+sr)^2g_s^2\Big(\langle
\bar{d}d\rangle^2+\langle \bar{q}q\rangle^2+2\langle
\bar{s}s\rangle^2\Big)}{2^{13}\cdot3^5\pi^6r^2}\Bigg\}\Theta[L],
\label{rho1}
\end{eqnarray}

and for the structure $\slashed q g_{\mu\nu}$:

\begin{eqnarray}
\rho^{p,\mathrm{pert.}}_{\frac{3}{2}}&=&\int\limits_{0}^{1} dx 
\frac{x^4(m_c^2+sr)^4\Big(-30m_s^2r(-4+r)+11x(m_c^2+sr)(-5+r)\Big)}
{2^{20}\cdot3^2\cdot5^2\pi^8r^4}\Theta[L],
\nonumber \\
\rho^{p}_{\frac{3}{2},3}&=&\int\limits_{0}^{1} dx \frac{m_sx^3(m_c^2+sr)^3\Big(40\langle
\bar{q}q\rangle-10\langle \bar{d}d\rangle(-3+r)+13\langle
\bar{s}s\rangle(-3+r)\Big)}{2^{15}\cdot3^2\pi^6r^2}\Theta[L],
\nonumber \\
\rho^{p}_{\frac{3}{2},4}&=&\int\limits_{0}^{1} dx \langle \frac{\alpha_s GG}{\pi}\rangle
\frac{x^2(m_c^2+sr)}{2^{19}\cdot3^3\cdot5\pi^6r^4}\Big[m_c^4x(-900+2215x
-1696x^2+326x^3)-5sr^3\Big(sx(-180+263x-63x^2)
\nonumber \\
&+&12m_s^2(-3+x^2)\Big)+
m_c^2r\Big(-12m_s^2(15-15x-10x^2+6x^3)+sx(-1800+4430x-3326x^2+641x^3)\Big)\Big]
\Theta[L],
\nonumber \\
\rho^{p}_{\frac{3}{2},5}&=&\int\limits_{0}^{1} dx \frac{m_sx^2(m_c^2+sr)^2m_0^2\Big(15\langle
\bar{d}d\rangle(-2+r)-14\langle \bar{s}s\rangle(-2+r)-45\langle
\bar{q}q\rangle\Big)}{2^{14}\cdot3^2\pi^6r}\Theta[L],
\nonumber \\
\rho^{p}_{\frac{3}{2},6}&=&\int\limits_{0}^{1} dx \Bigg\{\frac{x^2(m_c^2+sr)^2\Big[\langle
\bar{d}d\rangle\Big(3\langle \bar{q}q\rangle+10\langle
\bar{s}s\rangle(-2+r)\Big)-\langle \bar{s}s\rangle\Big(30\langle
\bar{q}q\rangle+\langle
\bar{s}s\rangle(-2+r)\Big)\Big]}{2^{11}\cdot3^2\pi^4r}
\nonumber \\
&+&\frac{11g_s^2\Big(\langle \bar{d}d\rangle^2+\langle
\bar{q}q\rangle^2+2\langle
\bar{s}s\rangle^2\Big)(m_c^2+sr)^2x^2(-2+r)}{2^{13}\cdot3^5\pi^6r}
\Bigg\}\Theta[L],
 \label{rho2}
\end{eqnarray}

where $\Theta[L]$ is the step function and
\begin{eqnarray}
L&=&-m_c^2x+sxr,
\nonumber \\
r&=&-1+x.
\end{eqnarray}
\end{widetext}
%


\end{document}